\documentclass{cjpsuppl}
\usepackage{epsfig }

\begin{document}
 \title{Analytic properties
of unitarization schemes}

 \authorii{J.-R. Cudell\footnote{e-mail: JR.Cudell@ulg.ac.be} }
 \addressii{Institut de Physique, B\^at. B5a, Universit\'e de Li\`ege, Sart Tilman, B4000  Li\`ege, Belgium}
 \authori{O.\,V. Selyugin \footnote{ e-mail: selugin@theor.jinr.ru}}
   \addressi{BLTPh, Joint Institute for Nuclear Research, Dubna, Russia}
 \authoriii{}
 \headtitle{Analytic properties
                     \ldots}
 \headauthor{O.V. Selyugin}
 \lastevenhead{J.-R. Cudell, \,  O.\,V. Selyugin : Analytic properties \ldots}
 \pacs{62.20}
 \keywords{high energies, unitarity, non-linear equations,
 saturation}
 \refnum{}
 \daterec{} 
 \suppl{?}  \year{2005} \setcounter{page}{1}
 \maketitle

 \begin{abstract}
  The analytic properties of the elastic hadron scattering
  amplitude are examined in the impact
  parameter representation at high energies.
  Different unitarization procedures and the corresponding non-linear
  equations are presented. Several unitarisation
  schemes are presented. They lead to almost identical results at
  the LHC.
\end{abstract}

 \section{Introduction}

Saturation is now  a very popular term, but it has a very
wide meaning. It includes shadowing and antishadowing processes,
the saturation of the Froissart-Martin bound \cite{frois},
gluon merging at small $x$ and geometrical scaling,
unitarization effects, etc. In  the $S$-matrix language,  saturation
means that we reach the maximum possible scattering. This happens
for a specific value of the impact parameter,
and is directly connected with the
unitary property of the scattering amplitude. It is that last meaning
that we shall  use  in our work.

The most important results on the energy dependence of diffractive
hadronic scattering were obtained from first principles
(analyticity, unitarity and Lorentz invariance), which lead to
specific analytic forms for the scattering amplitude as a function
of its kinematical parameters $-$ $s$, $t$, and $u$.
Analytic $S$-matrix theory relates the high-energy behaviour of
hadronic scattering to the singularities of the scattering amplitude
in the complex angular momentum plane.
One of the important theorems is the Froissart-Martin bound,
which states that the high-energy cross section for the
scattering of hadrons is limited by
 \begin{eqnarray}
 \sigma_{tot}^{max} = \frac{2 \pi}{\mu^2} \log^2 (\frac{s}{s_0}) ,
\label{FM}
\end{eqnarray}
where $s_0$ is a scale factor and $\mu$ the lightest hadron mass ({\it i.e.} the pion
mass). As the coefficient in front of the logarithm is very large, ``saturation of the
Froissart-Martin bound'' usually refers to an energy dependence of the total  cross
section rising as $\log^2 s$ rather than to the actual limit for the total cross section.

Experimental data reveal that total cross sections
grow with energy. This means that the leading contribution in the
high-energy limit is given by the rightmost singularity in the complex-$j$
plane, the pomeron,
with intercept exceeding unity. In the framework of perturbative QCD,
the intercept is expected to exceed unity by an amount
proportional to $\alpha_s$ \cite{lipatov1}. At leading-log $s$, one obtains
a rightmost singularity at $J-1 \  = \ 12 \ \log 2({\alpha_s }/{\pi})$.
Such a singularity $-$ the hard pomeron $-$ seems to be present in
inelastic diffractive
processes \cite{lnd-hp} and may be present in soft scattering,
where it appears as
a simple pole with intercept $\alpha_H\approx 1.4$
for energies smaller than 100 GeV \cite{cs6,mrt}.
In this case, the Froissart-Martin bound is soon violated, and the
BDL-regime may appear at relatively low energies. The effect of
saturation on the growth of the total cross section is however far
from clear, because it involves
processes at very small $x$ and non-perturbative effects.

Saturation  in the framework of perturbative QCD may be connected
with the growth of the gluon density at small $x$, which must be
bounded by non-linear effects.  In principle,
the gluon density may have a maximum which will  be reached
before the BDL (see {\it e.g.} \cite{levin-gd}).
In fact, the non-linear corrections added to the BFKL equation
lead to a saturation regime at relatively large impact parameters, and
are not directly related with the BDL, which first occurs at small impact parameters.
  One must note that, even in perturbative QCD, the description of saturation
at large impact parameters is problematic because of non-perturbative effects,
connected with confinement \cite{kovner-cf}.

\section{Unitarity: schemes and equations}

Unitarity of the scattering matrix
\begin{eqnarray}
  SS^{+} \ = \ 1.
\end{eqnarray}
is most suitably studied in
the impact parameter representation  \cite{t-m}
as it is equivalent (at high energy)
to a decomposition in partial-wave amplitudes.

We thus take the scattering amplitude in the impact parameter representation
\begin{eqnarray}
  T(s,t) =  \int_{0}^{\infty} \ b db J_{0}(b \Delta)
    G(b,s).
\end{eqnarray}
  with
\begin{eqnarray}
    Im G(b,s) \ \leq \ 1.
\end{eqnarray}
 and
\begin{eqnarray}
  Im G(s,b) = [Im (G(s,b)]^2 + [Re G(s,b)]^2 .
\end{eqnarray}

As energy grows, the scattering amplitude in the impact parameter
representation can exceed the unitarity bound for a range of
impact parameters $b<b_i$.
To restore unitarity, there are several different
schemes. Two of them are based on the solution of the unitarity equation
\cite{Pred-book}.
The first and most popular scheme seems to be the eikonal representation.
\begin{eqnarray}
  T(s,t) = i \int_{0}^{\infty} \ b db J_{0}(b q)
  \left(1\ - \ \exp(-  \chi(s,b)\right).
\end{eqnarray}
where the eikonal function $\chi(s,b)$, which in the non-relativistic
regime is equal to the scattering
potential, is often taken as the Born scattering
amplitude.

\begin{figure}[!ht]
\epsfysize=40mm \centerline{
\epsfbox{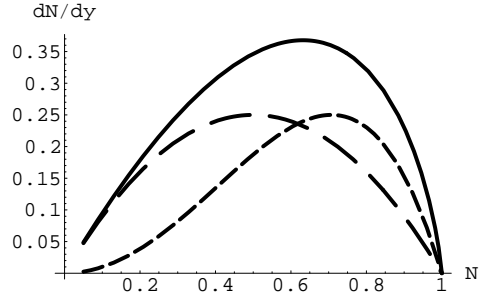}} \vspace{-0.5cm}
\caption{The derivative of the non-linear equations corresponding to
  the eikonal (hard line), the rescaled $U$-matrix (dashed line) and the
  hyperbolic tangent (long dashed line) unitarization schemes.
  }
\label{Fig_1b}
\end{figure}

The second unitarization scheme which we shall consider
is the $U$-matrix scheme
\cite{umat,savrin}, which we rescale so that
the unitarized amplitude becomes asymptotically
$i$ as $s\rightarrow\infty$:
\begin{eqnarray}
T(s,t) =  \int_{0}^{\infty} \ b db J_{0}(b q)
\left( \frac{ \chi(s,b)}{1-i \chi(s,b)})\right).
\end{eqnarray}
with $t=- q^{2}$.
If we assume that the Born scattering amplitude factorises as
   $f(b) s^{1+\Delta}$,
we can rewrite this scheme as the solution of the non-linear equation
\begin{eqnarray}
 \frac{dN}{dy} =\Delta N [1-N]) . \label{nl-n}
\end{eqnarray}
  We find that eikonal representation also corresponds the non-linear equation
  like above \cite{cs7}.
\begin{eqnarray}
 \frac{dN}{dy} = -\Delta \log(1-N) (1-N) . \label{nl-ln1}
\end{eqnarray}
We can also obtain the above schemes by assuming
that $dN/dN_{Born}\ = \ D(N)$, in which case
\begin{eqnarray}
 \frac{dN}{dy} = \frac{dN_{Born}}{dy} D(N) . \label{nl-ln2}
\end{eqnarray}

 We can also easily obtain new unitarization schemes \cite{cs8}, such as
\begin{eqnarray}
  T(s,t) =  \int_{0}^{\infty} \ b db J_{0}(b q)
   \ \tanh[ \chi(s,b)]  ).
\end{eqnarray}
which corresponds to the non-linear equation in the form
\begin{eqnarray}
 \frac{dN}{dy} = \frac{dN_{Born}}{dy} (1 - N^2) . \label{nl-ln3}
\end{eqnarray}

At first sight, these three schemes are quite different.
However we can represent them in one form
\begin{eqnarray}
  T(s,t) = i \int_{0}^{\infty} \ b db J_{0}(b q)
  \left(1\ - \ F( \chi(s,b)) \right).\label{common}
\end{eqnarray}

In order for Eq.~(\ref{common}) to produce a unitary amplitude, it
is sufficient to impose the following properties on $F$:\\
 a) it must be an analytic function of its variable $s$ and $b$,
    and satisfy crossing symmetry, \\
 b) it must not exceed unity, \\
 c) when $\chi(s,b) \ \rightarrow \ 0$,  $F \rightarrow \ 1$, \\
 d) when $\chi(s,b) \ \rightarrow \ \infty $, $F \rightarrow \ 0$, \\
e) at large $b$, it must fall faster than a power.

 The functions $F(s,b)$ of the above unitarization schemes
  satisfy such conditions. For the eikonal it has the standard form
\begin{eqnarray}
  F(s,b) \ = \ \exp[-  \chi(s,b)],
\end{eqnarray}
 for the rescaled $U$-matrix
\begin{eqnarray}
  F(s,b) \ = \
   \frac{1}{1+ \chi(s,b)} ,
\end{eqnarray}
and for hyperbolic tangent representation
\begin{eqnarray}
  F(s,b)  \ = \
   \frac{2 \exp(- 2 \chi(s,b)}{1+ \exp(- 2 \chi(s,b)} ,
\end{eqnarray}
 The differences between these functions is
  connected with different form of the saturation regime.


  In the following, we shall assume that
  the scattering
amplitude in the impact parameter representation can be factorised
 $ \chi(s,b) \sim h(s) f(b)$, and that
the energy dependence of $h(s)$ is
a power
\begin{eqnarray}
h(s) \sim s^{\Delta}.
\end{eqnarray}

We find that the energy dependence of the imaginary part of the amplitude and
hence of the total cross section depends on the form of
$f(b)$, {\it i.e.} on the $s$
and $t$ dependence of the slope of the elastic scattering amplitude.

 In Fig. 1 the process of saturation of the unitarity bound in the case
  of the three unitarization schemes is presented.
  It is clear that despite the slight difference in the growth
  the overlapping function with energy, the profile at the LHC will be
  similar for all these three unitarization schemes (see Fig. 2).

\begin{figure}
\vskip -0.5cm \epsfysize=40mm \centerline{
\epsfbox{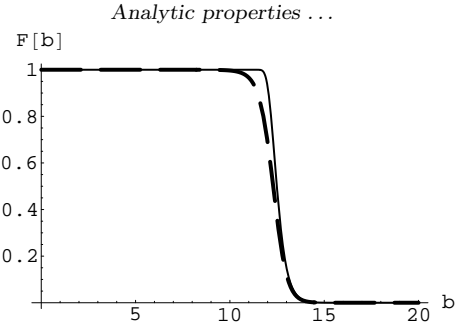}} \vspace{-0.5cm}
\caption{$G(s_{LHC},b)$, for the eikonal (hard line
 and the rescaled $U$-matrix (long-dashed line)
 forms at super high energies.
  }
\label{Fig_1c}
\end{figure}

   The overlapping function is complex,
   because it must be crossing
   symmetric. In that case, $s$ can be replaced by $s \ \exp(- i \ \pi/2) $.
   At very low energies, before the first inelastic channel opens,
   the total cross section is equal to the elastic cross section and
   the overlapping function has a large real part.
   When the energy grows and the overlapping function reaches the Black
   Disk Limit (BDL) (see, for example \cite{str-enk,bart}), the elastic
   cross section will be one half of the total cross section.
   Then the real part of the overlapping function
    disappears at small impact parameters. This has a small impact
    on the behavior of the total cross section, as shown in
Fig. 3, where we compare the values
   of $\sigma_{tot}$ in the case of a purely imaginary eikonal
   and of a complex eikonal.


\begin{figure}[!ht]
\vskip -0.5cm \epsfysize=60mm \centerline{
\epsfbox{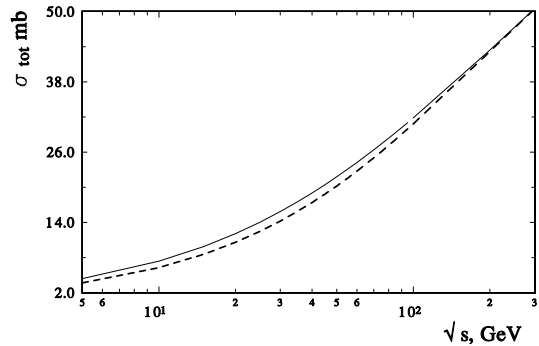}} \vspace{-0.5cm}
\caption{
  The total cross sections of proton-proton scattering
  are calculated for a purely imaginary eikonal (plain line) and for a complex eikonal (dashed line).
  }
\label{Fig_1}
\end{figure}

\section{Conclusion}
  We have shown that the impact parameter is the natural
variable to study unitarity of the scattering amplitude.
In particular, the unitarity bound leads to the Black
Disk Limit at high energies.
The corresponding saturation phenomena may be connected
with the saturation of the partonic density
   of the interacting hadrons. As a step in that direction,
we showed that
the usual unitarization schemes are in one-to-one
correspondence with nonlinear equations.
 Such an approach can be used
 to build new unitarization schemes
  and may also shed some light on the physical processes
underlying the saturation regime.
 In the presence of a hard Pomeron the saturation effects
 can change the behavior of some features of
  the cross sections already at LHC energies.
  In the unitarisation schemes considered here,
  non-linear effects appear before the BDL and lead to
 an acceptable growth of the total cross sections.
 Saturation then leads to a relative growth of the contribution
 of  peripheral interactions,
  and to changes in energy dependence of the differential cross
  sections  for moderate values of the  momentum transfer.
  Such effects in the differential elastic cross
  section can be  discovered at the LHC. \\

{\it Acknowledgments} We thank Jan Fisher
   for a stimulating discussion. O.V.S. acknowledges the support
  of FRNS (Belgium) for visits to
  the University of Li\`ege where part of this work was done.


\begin{thebibliography}{9}
\bibitem{frois} M. Froissart, Phys. Rev. {\bf 123} (1961) 1053;
  A. Martin, Nuovo Cimento {\bf A 42} (1965) 930.

\bibitem{lipatov1} L.N. Lipatov, Sov. J. Nucl. Phys. {\bf 23} (1976) 338;
 E.A. Kuraev, L.N. Lipatov, and V.S. Fadin,
 Sov. Phys. JETP {\bf 45} (1977) 199;
 I.I. Balitsky and L.N. Lipatov, Sov. J. Nucl. Phys. {\bf 28} (1978) 822.






\bibitem{lnd-hp}
S.~Donnachie, G.~Dosch, O.~Nachtmann and P.~Landshoff,
``Pomeron Physics And QCD,''
Cambridge: Cambridge University Press (2002)
(Cambridge monographs on particle physics, nuclear physics and cosmology. 19).

\bibitem{cs6}   J.-R. Cudell, A. Lengyel, E. Martynov, O.V. Selyugin,
       Nucl. Phys. A {\bf 755}, p. 587-590 [hep-ph/0501288].

\bibitem{mrt}J.-R. Cudell, A. Lengyel, E. Martynov and O.V. Selyugin,
                    {\it Phys. Lett.} {\bf B 587} (2004) 78;
       {\it Nucl.Phys.}  {\bf A 755} (2006) 587 [hep-ph/0501288].

\bibitem{levin-gd}  E. Levin, Nucl. Phys. {\bf A 763} (2005) 140.

\bibitem{kovner-cf} A. Kovner, Lectures on XLV Cracow School of Theoretical Physics,
        Zakopane, June (2005) , hep-ph/0508232.


\bibitem{t-m} K.A. Ter-Martirosyan,  Sov. ZhETF Pisma {\bf 15} (1972) 519;
   A.B. Kaidalov, L.A. Ponamarev, K.A. Ter-Martirosyan,
   Sov. J. Part. Nucl.  {\bf 44} (1986) 468.
\bibitem{Pred-book}    V. Barone and E. Predazzi,
  in book {\it "High-energy particle diffraction"}, Springer, New York, (2002).


\bibitem{umat} A.A. Logunov, V.I. Savrin, N.E. Tyurin, and O.A. Khrustalev,
  Theor. Mat. Fiz. {\bf 6} (1971) 157.

\bibitem{savrin} V.I. Savrin, N.E. Tyurin,  O.A. Chrustalev,
          Fiz. Elem. Chast. At Yadra  {\bf 7} (1976) 21.


\bibitem{cs7}   J.-R. Cudell, O.V. Selyugin,
          Nucl. Phys.  (Proc. Suppl.) {\bf B 146} (2005) 185 [hep-ph/0412338].

\bibitem{cs8} J.-R. Cudell, O.V. Selyugin,
        Czech. J. Phys. {\bf 55}  (2005) A235.


\bibitem{str-enk} P. Desgrolard,  L.L. Jenkovszky, B.V. Struminsky,
     Yad. Fiz. {\bf 63} (2000) 962.

\bibitem{bart}
J.~Bartels, E.~Gotsman, E.~Levin, M.~Lublinsky and U.~Maor,
Phys.\ Lett.\ B {\bf 556} (2003) 114
[arXiv:hep-ph/0212284].










\end{thebibliography}
\end{document}